\documentclass[useAMS,usenatbib]{mn2e}
\input psfig.sty

\title[Cosmological implications of APM 08279+5255, an old quasar at $z = 3.91$]{Cosmological implications
of APM 08279+5255, an old quasar at $z = 3.91$}
\author[J. S. Alcaniz, J. A. S. Lima and J. V. Cunha]{J. S. Alcaniz$^{1}$\thanks{E-mail:
alcaniz@astro.washington.edu} J. A. S. Lima$^{2}$\thanks{E-mail:
limajas@dfte.ufrn.br} J. V. Cunha$^{2}$\thanks{E-mail:
jvital@dfte.ufrn.br}\\ $^{1}$Astronomy Department, University of
Washington, Seattle, Washington, 98195-1580, USA\\
$^{2}$Departamento de F\'{\i}sica, Universidade Federal do Rio
Grande do Norte, C.P. 1641, 59072-970 Natal, RN, Brazil}
\begin{document}

\date{Accepted ; Received }

\pagerange{\pageref{firstpage}--\pageref{lastpage}} \pubyear{2002}

\maketitle

\label{firstpage}

\begin{abstract}
The existence of old high-redshift objects provides an important
tool for constraining the expanding age of the Universe and the
formation epoch of the first objects. In a recent paper, Hasinger
{\it et al.} (2002) reported the discovery of the quasar APM
08279+5255 at redshift $z=3.91$ with an extremely high iron
abundance, and estimated age of 2 - 3Gyr. By assuming the lower
limit for this age estimate and the latest measurements of the
Hubble parameter as given by the HST key project, we study some
cosmological implications from the existence of this object. In
particular, we derive new limits on the dark matter and vacuum
energy contribution. Our analysis is also extended to quintessence
scenarios in which the dark energy is parameterized by a smooth
component with an equation of state $p_x = \omega_x \rho_x$
($-1\leq \omega_x < 0$). For flat models with a relic cosmological
constant we show that the vacuum energy density parameter is
constrained to be $\Omega_\Lambda \geq 0.78$, a result that is
marginally compatible with recent observations from type Ia
supernovae (SNe Ia) and cosmic microwave background (CMB). For
quintessence scenarios the same analysis restricts the cosmic
parameter to $\omega_x \leq -0.22$. Limits on a possible first
epoch of quasar formation are also briefly discussed. The
existence of this object pushes the formation era back to
extremely high redshifts.
\end{abstract}


\section{Introduction}

An impressive convergence of recent observational results seems to
rule out with great confidence a large class of cold dark matter
(CDM) universes. To reconcile these observational evidence with
theory, cosmologists have proposed more general models whose the
basic ingredient is a negative-pressure dark component. The
existence of such a ``dark energy" not only explains the
accelerated expansion of the Universe observed from luminosity
distance measurements (Perlmutter {\it et al.} 1999; Riess {\it et
al.} 1998) but also reconciles the inflationary flatness
prediction ($\Omega_{\rm{Total}} = 1$) with the dynamical
estimates of the quantity of matter in the Universe which
consistently point to $\Omega_{\rm{m}} = 0.3 \pm 0.1$ (Calberg
{\it et al.} 1996). From a large variety
of independent techniques, the best-fit cosmological model with
the dark energy component represented by a cosmological constant
($\Lambda$CDM) has $\Omega_{\rm {m}} \sim 0.3$ and $\Omega_\Lambda
\sim 0.7$ (Peebles \& Ratra 2002; Padmanabhan 2002). If one assumes that the dark
energy is parameterized by a more general equation of state, say,
$p_x = \omega_x \rho_x$ with $-1 \leq \omega_x < 0$ (Turner \&
White 1997; Chiba {\it et al.} 1997) several analyses suggest
$\Omega_x \sim 0.7$ ($\Omega_{\rm {m}} \sim 0.3$) and $\omega_x <
-0.6$ as the best-fit quintessence scenario (see Kujat {\it et al.}
2002 and references therein). It is believed that with the new
generation of observational projects not only a more accurate
determination of the main cosmological parameters but also a
discrimination between general quintessence models and
$\Lambda$CDM scenarios ($\omega_x = -1$) will be possible.

On the other hand, if the presence of a dark energy component is
necessary in order to fit observational results with theoretical
predictions, its nature still remains a completely open
question giving rise to the so-called dark energy problem. In
this way, an important task nowadays in cosmology is to find new
methods or to revive old ones in order to quantify the amount of
dark energy present in the Universe, as well as to determine its
equation of state and/or its time dependence. In this
concern, the recent discovery of a 2-Gyr-old quasar at a redshift
of $z = 3.91$ is therefore a particularly interesting event. Its
existence is important to study the effect of a cosmological
constant or a quintessence component on the age of the Universe at
high-$z$ (Kennicutt Jr. 1996; Krauss 1997). In principle, together
with other recent age determinations of high-$z$ objects (Stockton
{\it et al.} 1995; Dunlop {\it et al.} 1996; Dunlop 1998; Yoshii
{\it et al.} 1998) a relevant statistical analysis may be
developed near future, thereby providing restrictive constraints
on any realistic cosmological model.

In the present work we study some cosmological implications from
the existence of the quasar APM 08279+5255. Although considering
the existence of several candidates to dark energy (Ratra and
Peebles 1988; Carvalho {\it et al.} 1992; Dev {\it et al.} 2002; Sahni and Shtanov 2002; Alcaniz 2002; Zhu
and Fujimoto 2002) we focus our attention to $\Lambda$CDM and
X-matter models (Turner and White 1997). In particular, we analyze
the constraints on the cosmological parameters $\Omega_\Lambda$
and $\omega_x$ from the age estimates of this object (Hasinger
{\it et al.} 2002; see also Komossa \& Hasinger 2002). The main
aim here is to convert the estimated lower bounds on the
dimensionless age parameter for this object to lower (upper)
limits on $\Omega_\Lambda$ ($\omega_x$). Some possible constraints
on the first epoch of quasar formation are also discussed. Our
approach is based on Alcaniz \& Lima (1999; 2001).

\section{Age-redshift test}

The age-redshift relation for a spatially flat, homogeneous, and
isotropic cosmologies with an extra smooth component ($p_x =
\omega_x \rho_x$) reads
\begin{eqnarray}
t_z & = &H_{o}^{-1}\int_{o}^{(1 + z)^{-1}}{dx \over
x\sqrt{\Omega_{\rm{m}}x^{-3} + \Omega_{x} x^{-3(1 + \omega_x)}}}
\nonumber \\ & &= H_{o}^{-1}f(\Omega_{\rm{m}}, \Omega_{x},
\omega_x, z) .
\end{eqnarray}
Note that for $\omega_x = -1$ the above expression reduces to the
well known result for $\Lambda$CDM scenarios ($\Omega_{x} \equiv
\Omega_\Lambda$) whereas for $\omega_x = 0$ the standard relation,
$t_z = \frac{2}{3}H_o^{-1}(1 + z)^{-3/2}$, is readily recovered. As can be easily seen from the above
equation limits on the cosmological parameters $\Omega_x$ and
$\omega_x$ can be readily obtained by fixing the product $H_ot_z$
from observations. Note also that such an age parameter depends
only on the product of the two quantities $H_o$ and $t_z$, which
are measured from completely independent methods. To clarify these
points, in what follows we briefly outline our main assumptions
for this analysis.

Following standard lines, we take for granted that the age of the
Universe at a given redshift is bigger than or at least equal to
the age of its oldest objects. As one may conclude, for
quintessence or $\Lambda$CDM scenarios, the comparison of these
two quantities implies a lower (upper) bound for $\Omega_{x}$
($\omega_x$), since the age of the Universe increases (decreases)
for larger values of this quantity. In order to quantify these
qualitative arguments, it is convenient to introduce the ratio
(Alcaniz \& Lima 1999)
\begin{equation}
\frac{t_z}{t_g} = \frac{f(\Omega_{\rm{m}}, \Omega_{x}, \omega,
z)}{H_o t_g} \geq 1,
\end{equation}
where $t_g$ is the age of an arbitrary object, say, a quasar or a
galaxy at a given redshift $z$ and $f(\Omega_{\rm{m}},\omega, z)$
is the dimensionless factor defined in Eq. (1). For each object,
the denominator of the above equation defines a dimensionless age
parameter $T_g = H_o t_g$. In particular, for the 2.0-Gyr-old
quasar at $z = 3.91$ yields $T_g = 2.0H_o$Gyr which, for the most
recent determinations of the Hubble parameter, $H_o = 72 \pm 8$
${\rm{km s^{-1} Mpc^{-1}}}$ (Freedman {\it et al.} 2001), takes
values on the interval $0.131 \leq T_g \leq 0.163$. It follows
that $T_g \geq 0.131$. Therefore, for a given value of $H_o$, only
models having an expanding age bigger than this value at $z =
3.91$ will be compatible with the existence of this object. In
particular, taking $\omega = 0$ (the standard flat case) in Eq.
(1), one obtains $t_z \leq 0.061$, which means that the
Einstein-de Sitter model is formally ruled out by this test
with great confidence (Komossa \& Hasinger 2002). In order to
assure the robustness of our analysis, we will adopt in our
computations the lower bound for the above mentioned value of the
Hubble parameter, i.e., $H_o = 64$ ${\rm{km s^{-1} Mpc^{-1}}}$.

\begin{figure}
\centerline{\psfig{figure=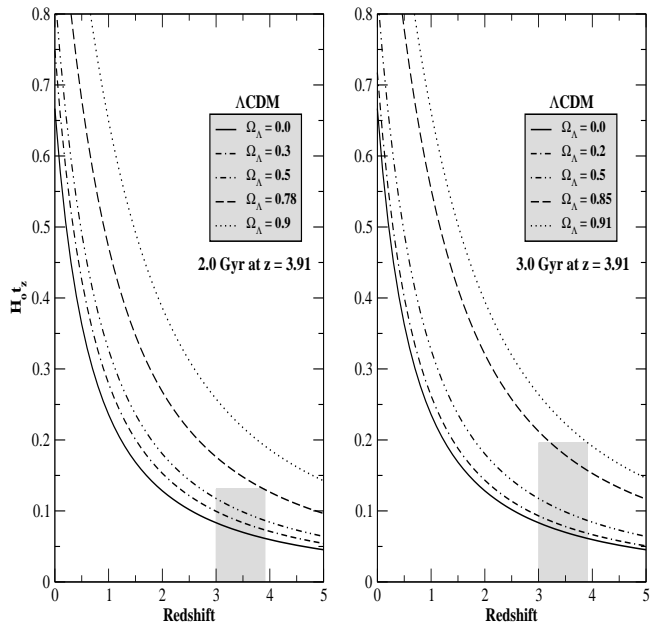,width=3.5truein,height=3.5truein,angle=-90}
\hskip 0.1in} \caption{a) Dimensionless age parameter as a
function of redshift for some values of $\Omega_{\Lambda}$. As
explained in the text, all curves crossing the shadowed area yield
an age parameter smaller than the minimal value required by the
quasar APM 08279+5255 reported by Hasinger {\it et al.} (2002). b)
The same as in Panel (a) for an age estimate of 3 Gyr.}
\end{figure}

To what extent does the Hasinger {\it et al.} (2002) result
provide new constraints on the cosmological parameters
$\Omega_{\Lambda}$ and $\omega_x$? To answer this question in
Figs. 1a and 1b we show the dimensionless age parameter $T_z =
H_ot_z$ as a function of the redshift for several values of
$\Omega_\Lambda$. The shadowed regions in the graphs were
determined from the minimal value of $T_g$. It means that any
curve crossing the rectangles yields an age parameter smaller than
the minimal value required by the presence of of the quasar APM
08279+5255. We see from Fig. 1a that by assuming an age estimate
of 2 Gyr the minimal value for the vacuum energy density is
$\Omega_\Lambda \geq 0.78$, which provides, for the above considered Hubble parameter interval, a minimal
total age of
$\sim$ 12. 8 - 16.0 Gyr. This value of $\Omega_\Lambda$ is only marginally in
accordance with most of the recent observational data which seem
to point out to a flat universe with $\Omega_\Lambda \simeq 0.7$ (see,
for example, Peebles \& Ratra (2002) for a recent review). We
recall at this point, in line with the arguments presented by
Komossa \& Hasinger (2002), that recent x-ray observations show an
Fe/O ratio for the quasar APM 08279+5255 that is compatible with
an age of 3 Gyr. In this case, Fig. 1b shows that the minimal
value of $\Omega_\Lambda$ required in order to make flat
$\Lambda$CDM scenarios compatible with the existence of this
object is $0.91$. Such a lower limit is much higher than the value
that can be inferred from an elementary combination of CMB
measurements pointing to $\Omega_{\rm{Total}} = 1.1 \pm 0.07$ (de
Bernardis {\it et al.} 2000; Jaffe {\it et al.} 2001) and
clustering estimates giving $\Omega_{\rm{m}} = 0.3 \pm 0.1$
(Calberg {\it et al.} 1996) or from a
rigorous statistical analysis involving many astrophysical
constraints (Harun-or-Rashid \& Roos 2001). Naturally, we do not
expect such results to be free of observational and/or theoretical
uncertainties (see, for example, Hamann \& Ferland 1993 and Chartas {\it et al.} 2002)\footnote{See also the
debate involving the age
estimates for the radio galaxy LBDS 53W069 (Bruzual \& Magris
1997; Yi {\it et al.} 2000; Nolan {\it et al.} 2000).}. However, if
independent analyses confirm such estimates the existence of this
quasar brings to light a new consistency problem between this
particular result and several others discussed recently in the
literature.

\begin{figure}
\centerline{\psfig{figure=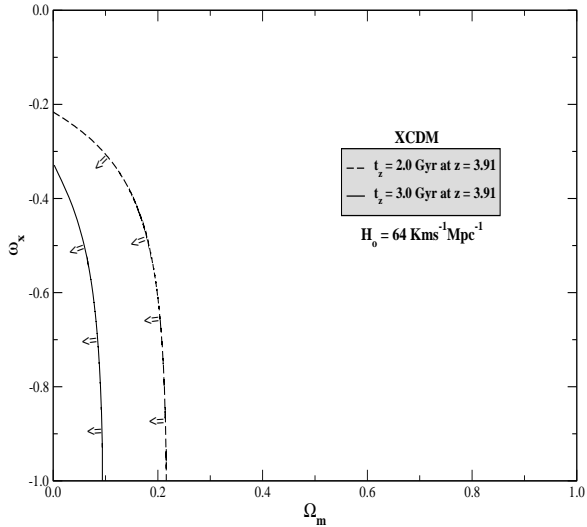,width=3.1truein,height=2.9truein,angle=-90}
\hskip 0.1in} \caption{Countours of a fixed age parameter $H_ot_z$
for the quasar APM 08279+5255. The solid curve corresponds to an
age estimate of 3 Gyr while the dashed one stands for 2 Gyr. For
each contour the arrows point to the allowed parameter space. We
see that for $\omega_x = -1$ the results of Figs. 1a and 1b are
recovered.}
\end{figure}

Figure 2 shows a similar analysis for quintessence scenarios. Now,
instead of plotting the age-redshift diagram for these models we
show the parameter space in which contours represent the minimal
value of the age parameter discussed above. As can be seen,
although restrictive constraints can be placed on the matter
density parameter (in line with the above discussion), the upper
limits on the quintessence equation of state are not so tight.
This happens basically because at this redshift the age parameter
is not a very sensitive function to the equation of state
parameter $\omega_x$. In particular, we found $\omega_x \leq
-0.22$ and $\Omega_{\rm{m}} \leq 0.22$ and $\omega_x \leq -0.33$
and $\Omega_{\rm{m}} \leq 0.09$ for an age estimate of 2 and 3
Gyr, respectively. Such bounds on $\omega_x$ are very similar to
that ones obtained from the age estimates of the radio galaxies
LBDS 53W069, LBDS 53W091 and 3C 65, although the limits on
$\Omega_{\rm{m}}$ are now much more restrictive (Lima \& Alcaniz
2000). The latter aspect is particularly noticeable to the
case of a cosmological constant ($\omega = -1$). From figure 2 we
see that for a $\Lambda$CDM Universe, the density parameter is
just the extreme value, $\Omega_{\rm{m}}=0.22$, when the lower limit of 2 Gyr is considered. However, by
taking 3 Gyr as the real age of the quasar
one finds the second extreme value, namely, $\Omega_{\rm{m}}=0.09$.
Both results disagree completely with the recent limits on
$\Omega_{\rm{m}}$ obtained from X-ray measurements of galaxy
clusters (Allen {\it et al.} 2002). The main results of this analysis
are presented in Table I.

\begin{figure}
\centerline{\psfig{figure=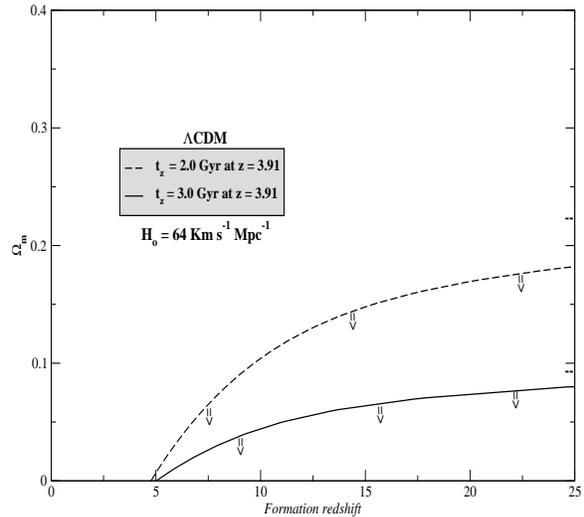,width=3.1truein,height=2.9truein,angle=-90}
\hskip 0.1in} \caption{ The $\Omega_m-z_f$ plane for
$\Lambda$CDM models. The contourns are fixed using the age
parameter $H_ot_z$ for the quasar APM 08279+5255. The dashed curve
corresponds to 2 Gyr while the solid one represents an age estimate
of 3 Gyr. For each contour the arrows delimit the allowed
parameter space. We see that the redshift formation increases with
the value of $\Omega_m$. However, the allowed region for each
curve falls below the interval ($\Omega_m=0.3 \pm 0.1$) inferred
from independent methods.}
\end{figure}

\begin{table}
\begin{center}
\begin{tabular}{rll} \hline \hline
\multicolumn{1}{c}{Age estimate}&
\multicolumn{1}{c}{$\Omega_{\Lambda}$}&
\multicolumn{1}{c}{$\omega_x$}\\ \hline 2.0 Gyr..........& $\geq
0.78$ & $\leq -0.22$ ($\Omega_{\rm{m}} \leq 0.22$)\\
3.0 Gyr..........& $\geq 0.91$ & $\leq -0.33$ ($\Omega_{\rm{m}}
\leq 0.09$)\\ \hline \hline
\end{tabular}
\caption{Limits to $\Omega_{\Lambda}$ and $\omega_x$}
\end{center}
\end{table}

\section{Implications on the epoch of quasar formation}

At this point we change the focus of our discussion to investigate
possible constraints on the epoch of quasar formation from the age
estimate of the quasar APM 08279+5255. In order to infer such
limits we do not consider in our computations the time necessary
for the quasar formation. In other words, the reasonable
assumption of an incubation time will be totally neglected in our
analysis. This means that any limit on the redshift formation
$z_f$ will be a conservative lower bound. For look-back time
calculations, such a hypothesis can be translated as (Alcaniz \&
Lima 2001)
\begin{eqnarray}
t_{z_{obs}} - t_{z_f} & = & H_o^{-1}\int_{(1 + z_f)^{-1}}^{(1 +
z_{obs})^{-1}}\frac{dx}{x\sqrt{\Omega_{\rm{m}}x^{-3} + \Omega_{x}
x^{-3(1 + \omega_x)}}} \nonumber \\ & & \geq t_g,
\end{eqnarray}
where the inequality signal comes from the fact that the Universe
is older than or at least has the same age of any observed
structure. Since this natural argument also holds for any time
interval, a finite value for the $z_f$ will contribute to make our
lower limits on the epoch of quasar formation even more
conservative. Naturally, models for which $z_f \rightarrow \infty$
are clearly incompatible with the existence of this particular
object, thereby being ruled out in a natural way.

In Fig. 3a we show the $z_f - \Omega_{\rm{m}}$ plane allowed by
the existence of the quasar APM 08279+5255 for $\Lambda$CDM
models. As before, two age estimates were assumed: 2 Gyr (dashed
curve) and 3 Gyr (solid curve). As should be physically expected,
since the effect of dark matter is decelerate the cosmic
expansion\footnote{It means that the look-back time between the
observed redshift $z_{obs}$ and $z_f$ is smaller for larger values
of $\Omega_{\rm{m}}$.}, the larger the contribution of
$\Omega_{\rm{m}}$ the larger the value of $z_f$ that is required
in order to account for the existence of this quasar within these
cosmological scenarios. In this way, the smallest value for the
formation redshift occurs for a completely empty universe
($\Omega_{\rm{m}} = 0$). From Figure 3 one may see that $z_f \geq
5$. For a low-density universe with $\Omega_{\rm{m}} = 0.2$ we
obtain $z_f \geq 40$. As expected from the previous analysis, $z_f
\rightarrow \infty$ when the matter density parameter approaches
to 0.22.

\section{conclusion}
The problem related to the lower limits for the total age of the
universe, as inferred from age estimates of globular clusters (at
$z=0$), was a recurrent problem in the development of physical
cosmology. In actual fact, it was a real source of progress for
cosmology ever since the Hubble discovery of the expanding
Universe.

In a similar vein, old high redshift objects may play an
equivalent role to the question related to the ultimate fate of
the Universe. Their age estimates provide a powerful technique
for constraining the basic cosmological parameters. In particular,
the so-called high redshift ``age crisis" is now becoming an
important complement to other independent cosmological tests. As
we have seen, the constraints on the $\Omega_{\rm{m}}$ - $\Omega_\Lambda$
plane based solely on the recent age estimates of the APM
08279+5255 quasar are very restrictive. Actually, the analysis
presented here is more restrictive than the earlier results
derived from age estimates of some old high redshift galaxies (Alcaniz \& Lima
1999). It should be stressed that the present constraints on the
formation redshift are indeed rather conservative since the lower
limit on the age of the APM quasar has been considered in all the
estimate and the possible incubation time has also been neglected
in the original work. Finally, these high limits on $z_f$ are also
in line with the current idea that old objects are not uncommom at
very large redshifts, say, at $z > 4$, and also reinforce the
interest on the observational search for quasars, galaxies and
other callapsed objects within the redshift interval $6 \leq z
\leq 10$, which nowadays delimits the so-called dark zone.

\section*{Acknowledgments}
The authors are grateful to Scott F. Anderson for helpful discussions. This work is supported by the Conselho
Nacional de Desenvolvimento
Cient\'{\i}fico e Tecnol\'{o}gico (CNPq - Brazil) and CNPq (62.0053/01-1-PADCT III/Milenio).

\end{document}